# Dynamic Voltage Restorer (DVR) For Protecting Hybrid Grids

Khodakhast Nasiriani[1], Mohsen Pasandi[2]

[1]Ph.D. student in Electrical Engineering Department, Marvdasht Azad University, Marvdasht, Iran

[2]Master's Student, Department of Electrical Engineering, Clausthal University of Technology, Germany

*Abstract*—Dynamic Voltage Restorer (DVR), for some important reasons, is utilized in Power Distribution for protecting loads especially critical loads against some power quality issues like sag, swell, harmonics and faults. Fast transient response for compensating voltage quality is an unavoidable factor. Different control topologies have been applied to this device. In this project, a DVR is deployed to preserve a critical load which is connected in parallel with grid and A wind turbine. Several situations of power quality issues have been considered. The simulation results of a proposed system which has been implemented in MATLAB/Simulink show how DVR is effective for protecting loads.

## I. Introduction

In line with the design and development of electronic power systems, electric power research institutions have defined the concepts of FACTS in transmission systems and Custom Power in distribution systems. Custom Power devices have the same physical arrangement as the FACTS arrangement, but their application and control are completely different with FACTS devices [1-3]. Therefore, the use of power electronic devices is introduced to improve the quality of power in distribution systems which called Custom Power. Custom Power Equipment is designed to improve the reliability and quality of power or specifically voltage in distribution networks [3-5].

An example of this equipment is the Dynamic Retrieval of Voltage (DVR), which has received a lot of attention in recent years and has been introduced as the best voltage recovery equipment. The factors that have created the voltage shortage is one of the issues of power quality as well as the relations related to its calculation and modification techniques in the distribution networks, are expressed mainly for several faults conditions such as three-phase balanced faults, two-phase unbalanced to ground fault and single-phase fault. The methods which are used to compensate it are often power-based electronic devices. Some of these solutions are using SMES, D-Statcom, UPQC, and so on [5-8].

Various types of custom power equipment and their tasks and applications which includes two categories are presented. The first group is those which do not need a DC power source and are used to change the network structure. This equipment is mainly used as static breakers in the systems, and they have the role of ordinary mechanical breakers. Generally, this equipment is a subset of solid-state breakers (SSBs) or fast-disconnected semiconductor circuits. The second category includes devices that come with a DC power source and is used for compensating purposes, the most widely used are DVR, D-Statcom and UPQC [9-12].

The DVR has been introduced as an appropriate device for protecting sensitive loads against disturbances in the distribution network and In terms of its application to correct the voltage shortage in the network, the voltage has been calculated by the DVR to compensate for the voltage deficiency based on the injection power [14].

Fault detection and determining of signal reference are two main parts of a control system in a DVR. Fault detection part is the first step which the voltage grid will be measured and analyzed base on the method which has been used. The harmonic and disturbances in voltage can be recognized. Several detection methods like peak measurements, RMS measurement, dq0 components, positive sequence, using Kalman filter for estimating phasor components or Fourier Transformation have been reported in papers. In the second part, the method for determining the reference signal of series injected voltage is related to Energy Storage Unit. Several controllers such as PI, fuzzy, Neural Networks has been used in DVR [11-15].

Renewable Energies such as wind energy have several merits and using of them is increasing significantly. Several aspects like protection, stability load management, and power quality should be considered for them. Power quality issue is a very important issue for them. Sometime during the fault condition, the wind system is separated from the grid. But by using DVR, the system can be protected without separation.

Forecasting is a decision-making technique employed by many to help in budgeting, planning, and estimating the future. In the simplest terms, Forecasting is an effort to predict future outcomes based on past events and management insight. Different types of Forecasting are widely used in business, finance, engineering, and other science. The time-series econometrics model, applying statistical methods to economic data, is commonly used in economics and business (see [16]). The time series model also used in electrical engineering to predict price, load, and wind speed (see [17]).

A solar system with P&O Maximum Power Point Tracking has been used in this research for hybrid system. Several essential aspects of the wind turbine should be investigated more. Forecasting of wind speed can play a significant role in the planning and operating of the wind turbine. The weak effect is another essential item that can be considered more [18]. DVR control strategies can be considered in three categories: open loop, closed-loop ant the improvement of the control strategy. Implementing of open loop is easy but it does not have enough precision. Therefore, closed-loop control can improve control precision and stability.

In this project, a DVR is used to protect a hybrid system included solar, and a constant speed wind turbine against voltage fluctuations. The DVR will keep the voltage on the busbar for load and wind system almost constant. The simulation results of a proposed system which has been implemented in MATLAB/Simulink show how DVR is effective for protecting loads.

## II.  POWER QUALITY

Two of the most important problems of power quality is voltage sag and voltage swell, which are of the most important challenges faced by power distribution companies in dealing with advanced industries, and about 80% of the quality problems of the system are included. Different Standards have different definitions for voltage quality. In the IEEE-1159-1995 standard, the voltage sag is defined as: "A sudden decrease in the effective value of a voltage of 10% to 90% at a nominal frequency and within a time interval of half cycle to 1 minute." In this definition, the voltages below 10% of the nominal voltage are considered as voltage interruption or voltage cottage. In the IEEE-1159-1995 standard, the voltage swell is defined as: "The sudden increase in the effective value of the voltage over 110% at the nominal frequency and in the period from half cycle to 1 minute." If voltage swell exceeds one minute, it is overvoltage and over-voltage of less than half a cycle is considered transient. In this definition, for both the maximum and minimum voltage, a threshold is considered. Figure 2.1 illustrates the definition of deficiency and voltage overvoltage in IEEE-1159-1995 [5].  Figure (1): Definition diagram for voltage sag and swell by IEEE-1159-1995 standard [3].

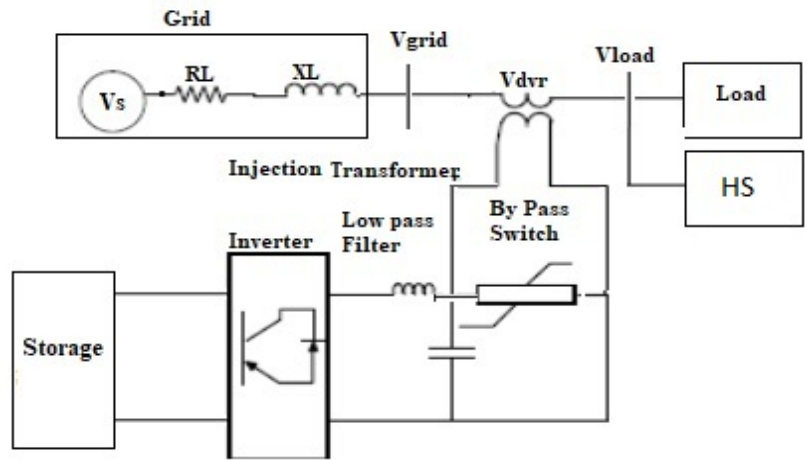

Fig 1: overview of the DVR

### III. DVR COMPONENTS

In this chapter, the DVR in terms of the structure of the circuit, its components and how it is designed and the factors which are effective in its design is described. Figure (1) shows the overall structure of the DVR, which includes the following units: Figure (2): Dynamic Voltage Restorer structure

*A. Energy Storage Unit*

For low voltage sag, load voltage magnitude just by injecting reactive power can be compensated. But, for compensating deep voltage sag, in addition to reactive power injections, the injection of active power is also required.

*B. Voltage Source Inverter (VSI)*

The main task of the Voltage Source Inverter (VSI) or inverter is to convert the DC voltage provided by the energy source to the AC voltage.

*C. Passive filters*

Low-Pass passive filters are used to convert the PWM pulse waveform into a sinusoidal waveform.

*D. Voltage injection Transformers*

The main task of the transformer is to increase the voltage supplied by the VSI to the desired level and isolate the DVR circuit from the distribution network.

*E. By-Pass Switch*

This switch is used to protect the inverter from high currents. When a fault occurs on the downside, By-pass switch will be active and as a result, DVR changes the situation to By-Pass and protect (VSI) from high current.

*F. DVR Control System*

In general, the DVR control system has the following functions:

- Error detection
- Calculate and determine the voltage required for compensation
- Production of trigger pulses for VSC and serial voltage injection

- Termination of triggered pulses when the error is resolved.
- Control the security Breakers and disconnect the DVR to the network

It is also possible to use a control system to change the DC-AC inverter to a rectifier (rectifier) mode for charging capacitors in a DC power link in the absence of an error [2],[4].

## IV. WIND TURBINE MODELING

Using wind energy started in 19th century. At the first it was so expensive. But today by improving the technology is getting cheaper and cheaper. It can be a good replacement for fuel resources. Several kinds of generators are utilizing in wind systems as following:

1. DC Generators
2. AC Synchronous Generators
3. AC Asynchronous Generators
4. Switched Reluctance Generators

The induction generators are categorized into two categories: fixed-speed induction generators (FSIGs) (Fig. 2) with squirrel cage rotors, and doubly fed induction generators (DFIGs) with wound rotors [19-20].

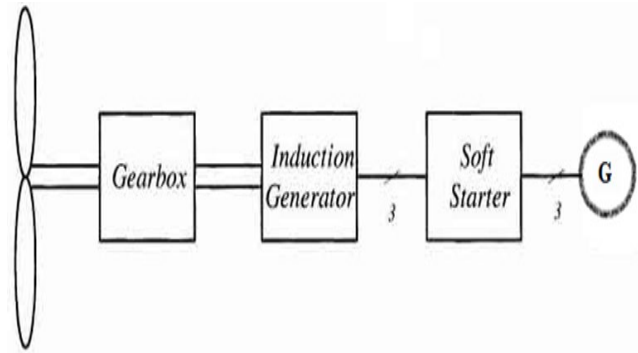

Fig 2. fixed-speed induction generators (FSIGs)

Based on the design, the rotor dimension, and other parameters the generated power can be different. The extracted power from wind turbine can change based on below equations [21-22]:

$$P_m = C_p P_w \qquad (1)$$

$$P_w = \frac{1}{2} \pi \rho R^2 v_w^3$$

$P_m$ is mechanical power, $P_w$ is wind power. $\rho$ air density, R is the radius of the turbine blades, $V_w$ wind speed, and $C\rho$ is Performance coefficient. In this research, an induction generator for wind generator is used. Below equations show how the generator is formulated. This model is shaped based on DQ rotating frame without zero component.

$$\begin{bmatrix} \lambda_{ds} \\ \lambda_{qs} \\ \lambda_{dr} \\ \lambda_{qr} \end{bmatrix} = \begin{bmatrix} L_{ls}+L_m & 0 & L_m & 0 \\ 0 & L_{ls}+L_m & 0 & L_m \\ L_m & 0 & L_{lr}+L_m & 0 \\ 0 & L_m & 0 & L_{lr}+L_m \end{bmatrix} \begin{bmatrix} i_{ds} \\ i_{qs} \\ i_{dr} \\ i_{qr} \end{bmatrix} \quad (2)$$

Below equations show that how the extended torque in the generator shaft and the angular velocity of the rotor are interrelated

$$J_R \frac{d\omega_r}{dt} = T_g - T_{em} \quad (3)$$

## 1-2) Solar Cell Modeling

PV cell simulation includes obtaining characteristic (P-V) and (I-V) curves. The purpose of this work is to adapt the curves Characteristics of the Simulation Model with the Characteristic Curve of the actual Cell under various environmental conditions. The most common approach is to use an equivalent electrical circuit which it is based on a diode model. Many models have been presented by several researchers which the simplest model, is Single diode model. This model includes an independent current source, which has been connected parallel with a diode that shown in figure (3). The mathematical equations which represent the performance of this model are expressed by [1].

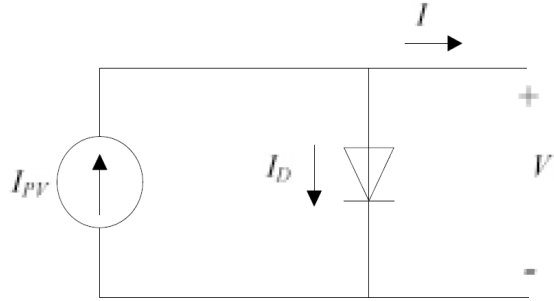

Figure 3. The ideal solar cell model

$$I = I_{pv} - I_D \quad (4)$$

$$I = I_{pv} - I_0 \left[ \exp\left[\frac{V}{\alpha V_T}\right] - 1 \right] \quad (5)$$

$$V_T = \frac{KT}{q} * nl * N_{cell} \quad (6)$$

In this project first method has been used. This algorithm perturbs the operating voltage to ensure maximum power. While there are several advanced and more optimized variants of this algorithm, a basic P&O MPPT algorithm is shown below.

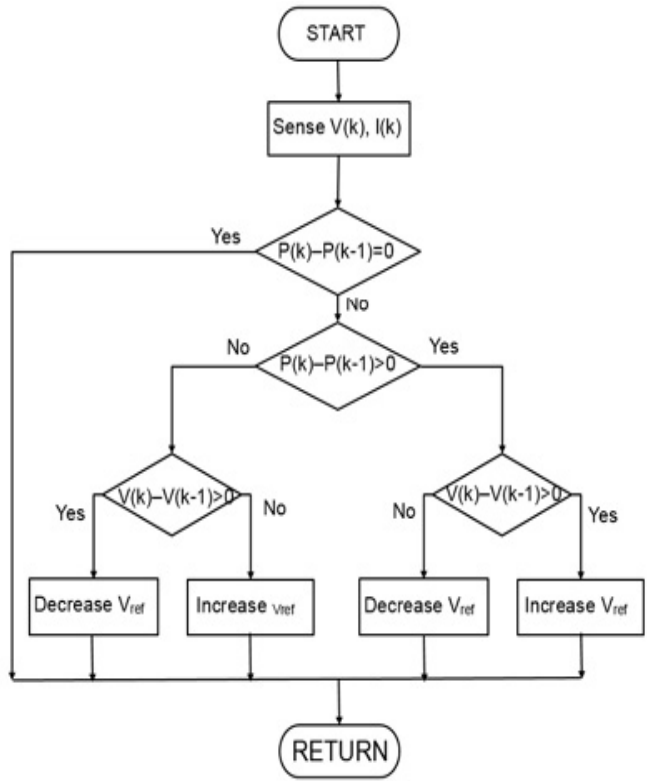

Figure 4. Perturbation and observation (P&O)

V. CONTROL SYSTEM

The below figure shows the circuit for the system.

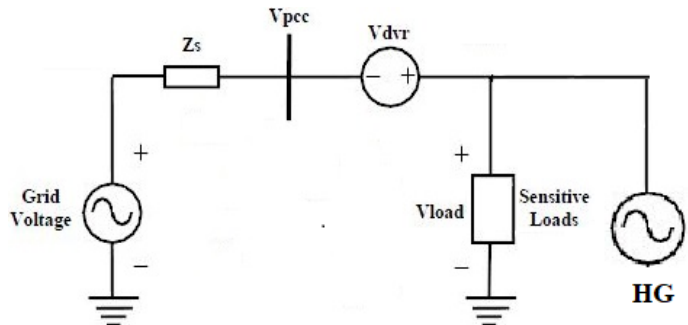

Fig 5. DVR circuit

Based on above figure, the following equation can be achieved.

$$V_{Load} = V_{grid} - Z_{th}I_L + V_{dvr} \qquad (7)$$

The Fig. (3) Gives a view of the used control strategy for this system. The $V_{\_load}$ is taken and given to a block to get fundamental positive sequence. The taken value based on per unit will be sent to be compared with the reference value which is 1pu. The Pi control will control the error to send the value to PWM. The output of PWM will be sent to the gate of VSC for controlling the system.

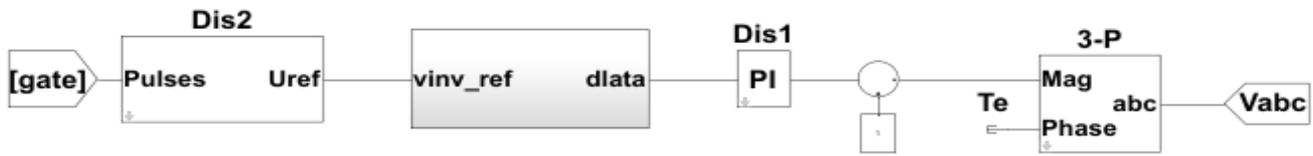

Fig 6. used control strategy

## VI. SIMULATION AND RESULTS

The below system is designed and simulated to protect critical load and wind turbine against several voltage quality issues. The simulations have been done in MATLAB/Simulink environment. Table (1) shows the characteristics of the studied system.

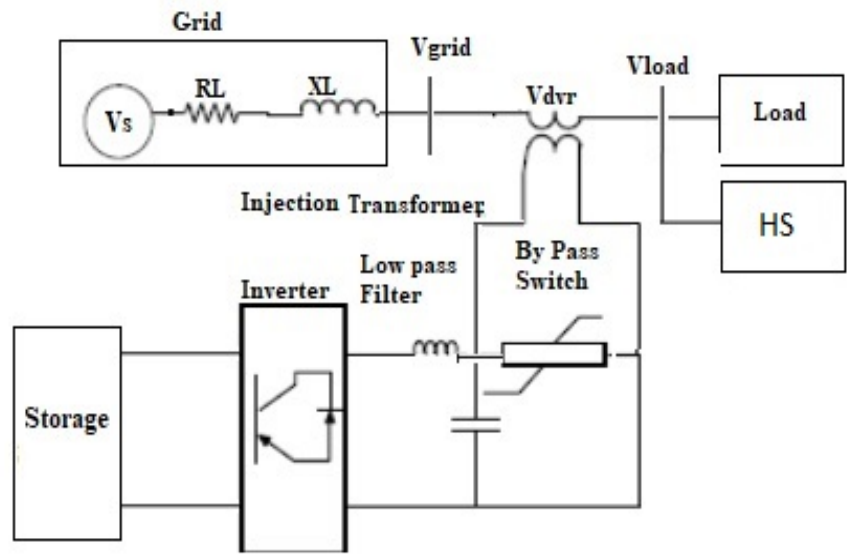

Fig 7. Simulated system in MATLAB

TABLE I
SYSTEM CHARACTERISTICS

| Voltage L_L | 20 KV |
|---|---|
| Frequency | 50 HZ |
| transformer | 20kv/400v Yg-Yg |
| Load Active Power | 20 KW |
| Load Reactive Power | 15 KAR |
| Wind turbine Voltage | 400 V |
| DVR filter | |
| L | 50mh |
| R | 0.005 oh |
| DC voltage DVR | 800 |

### A. Case 1

:
Fig (7) shows applying two different stages sage 20% and 50% sag to the grid. Fig shows the good performance of the design

DVR two different levels of sag are applied to the grid voltage. The simulation is designed for 1 sec. after 0.2 s a sag with 20% depth occurs. The second sags depth occurs after 0.4 sed and it will be finished after 0.6 second.

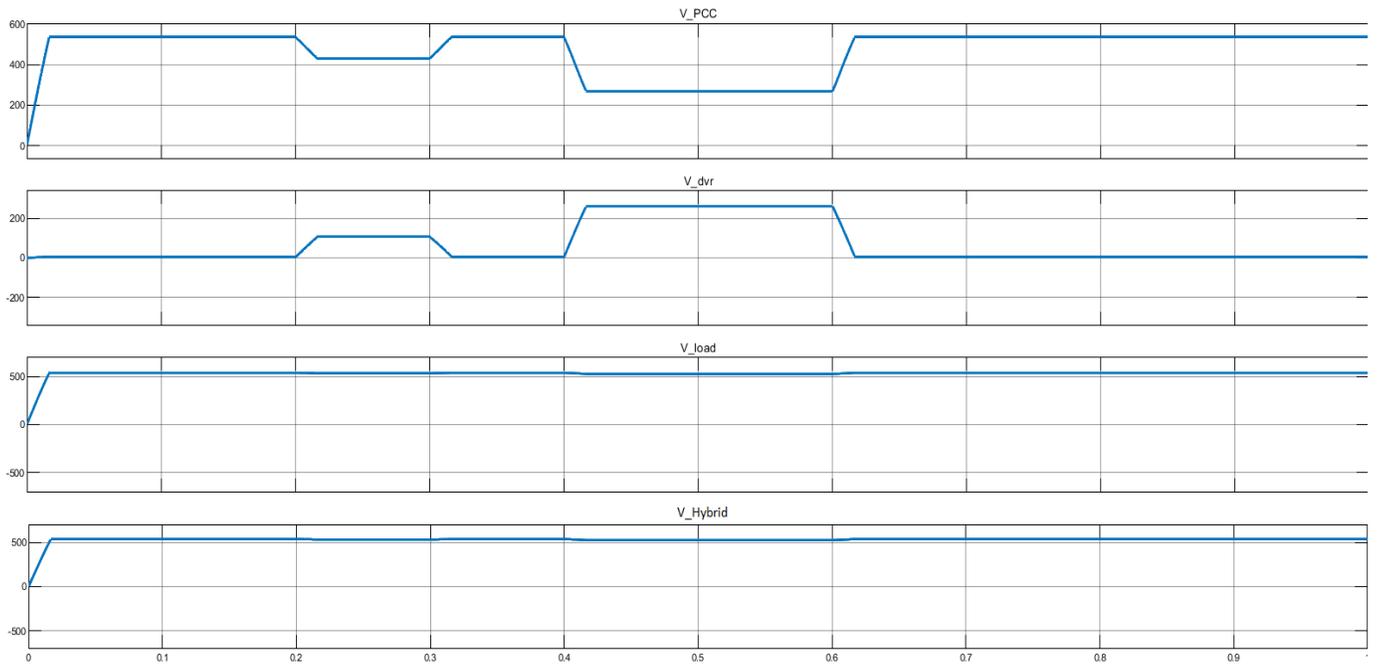

Fig 8. RMS voltage values for sag

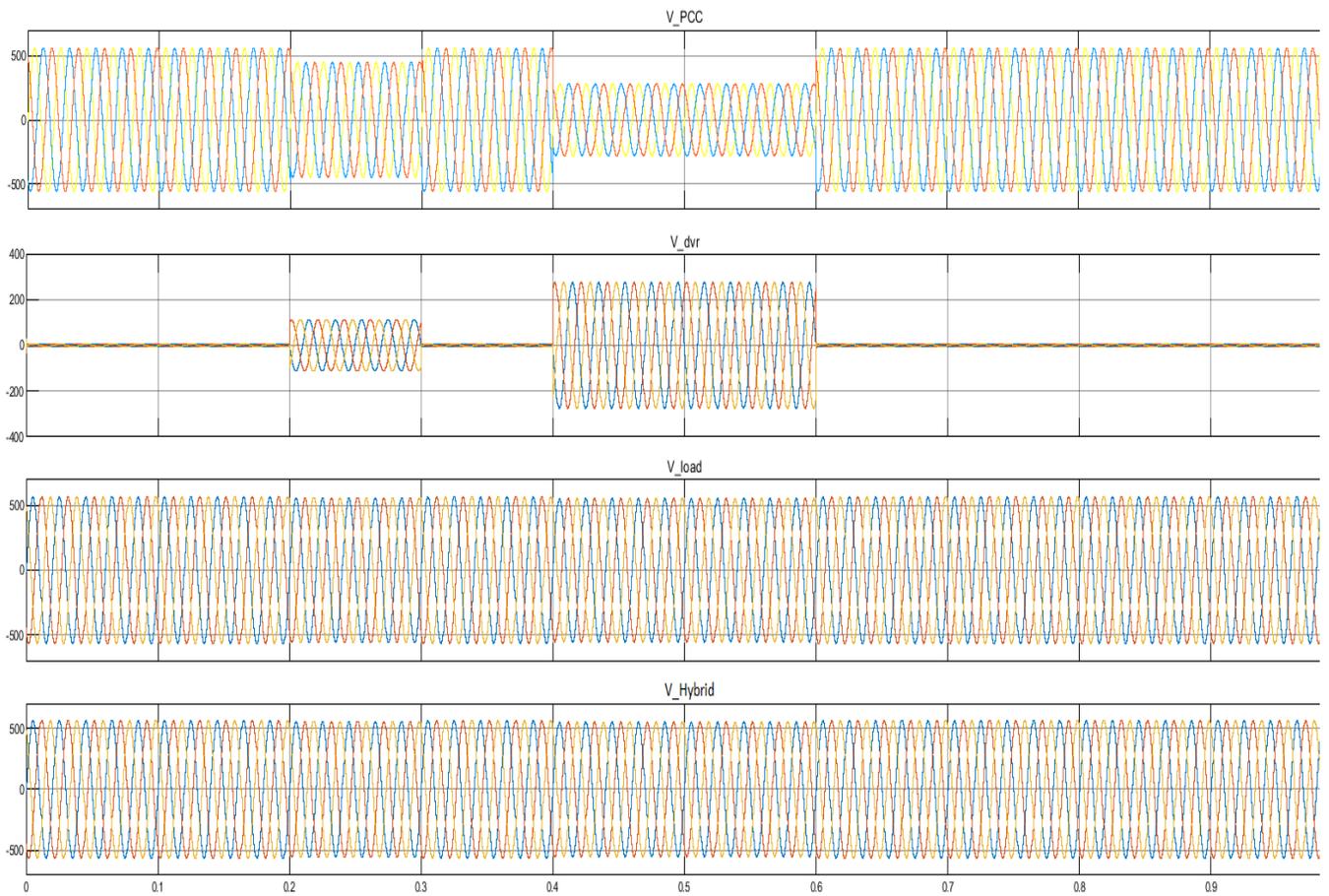

Fig 9. Applying two different stages sage 20% and 50% sag to the grid

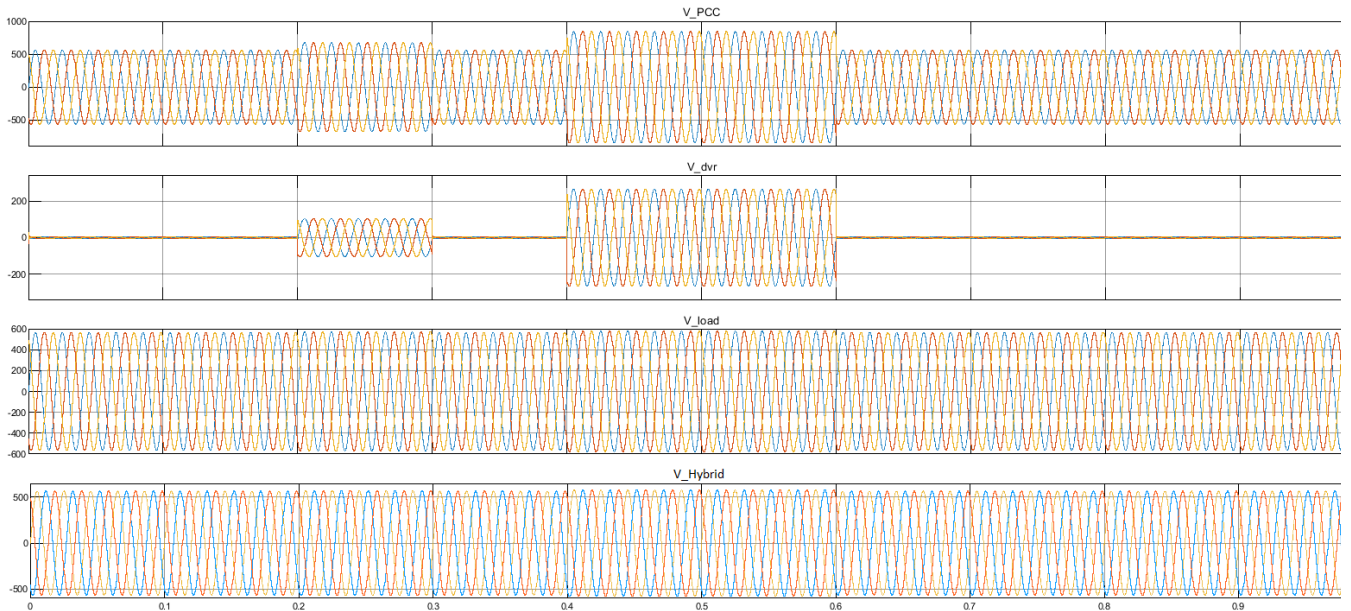

Fig 10. Applying two different stages sage 20% and 50% sag to the grid

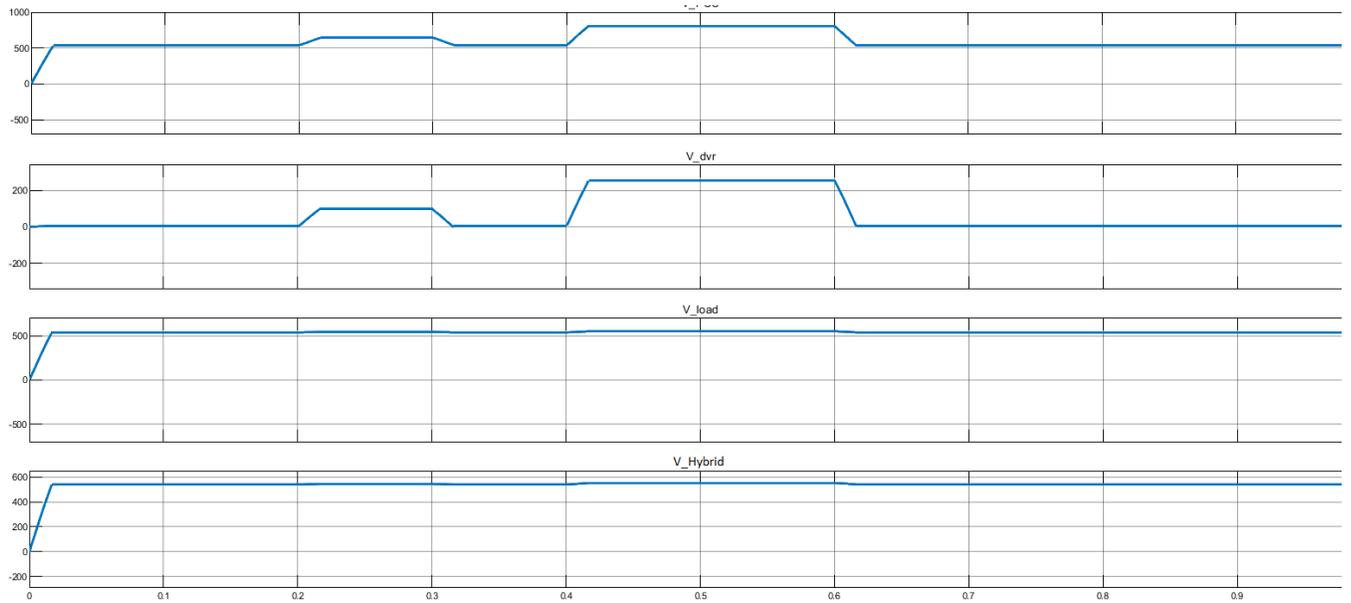

Fig 11. RMS voltage values for swell

Fig (6) presents a good view of $V_{rms}$ for grid, DVR, load, and wind. The good capability of DVR to remove the swell is depicted.

B. *Case 2:*

To show the great presentation of the modeled DVR two various heights of swell is picked for utility. The simulation is designed for 1 sec. after 0.2 s a swell with 20% depth occurs. The second swell depth occurs after 0.4 sec and it will be finished after 0.6 second (see fig 8). Fig (8) presents a good view of $V_{rms}$ for grid, DVR, load, and wind. The good capability of DVR to remove the swell is depicted.

VII. CONCLUSIONS

In this paper, we presented a topology for protecting a wind turbine against voltage sag and swell. The Dynamic Voltage Restore (DVR) as a series compensator was utilized for injecting proper voltage according to the grid voltage. The load voltage for sensitive load and wind farm stayed almost constant for studied cases. The effectiveness of proposed method was verified by simulation in the MATLAB/Simulink environment.